\documentclass[12pt]{article}

\usepackage[margin=1in]{geometry}
\usepackage{amsmath}
\usepackage{amssymb}
\usepackage{graphicx}
\usepackage{booktabs}
\usepackage{array}
\usepackage{longtable}
\usepackage{hyperref}
\usepackage{natbib}
\usepackage{setspace}
\usepackage{caption}
\usepackage{subcaption}
\usepackage{float}
\usepackage{authblk}
\usepackage{xcolor}
\usepackage{url}

\graphicspath{{figures/}}

\hypersetup{
    colorlinks=true,
    linkcolor=blue,
    citecolor=blue,
    urlcolor=blue
}

\title{\textbf{Evacuation destination choices during Hurricane Ian: A direct demand modeling approach}}

\author[a,*]{Alessandra Recalde}
\author[b]{Luyu Liu}
\author[a]{Xiaojian Zhang}
\author[a]{Sangung Park}
\author[a]{Shangkun Jiang}
\author[a]{Xilei Zhao}

\affil[a]{Department of Civil and Coastal Engineering, University of Florida, USA}
\affil[b]{Department of Geosciences, Auburn University, USA}

\date{}

\begin{document}

\maketitle

\noindent\textbf{*Corresponding author:} Department of Civil and Coastal Engineering, University of Florida, 1949 Stadium Rd, Gainesville, FL 32611, USA.\\
\noindent\textbf{E-mail address:} \href{mailto:arecalde@ufl.edu}{arecalde@ufl.edu} (A. Recalde).

\bigskip

\noindent\textbf{Keywords:} Hurricane evacuation; Direct demand; Mobile device location data; Social vulnerability; Hurricane Ian

\begin{abstract}
Hurricanes are causing unprecedented damage to the natural environment, infrastructure, and communities. Understanding evacuation behavior is essential for improving emergency preparedness. Past studies have relied on surveys and interviews, which are prone to recall bias. Additionally, they urge incorporating social vulnerability in evacuation research, emphasizing its impact on evacuation capability and destination choice. This study addresses these gaps by analyzing evacuation behavior using mobile device location data from Hurricane Ian, one of Florida's deadliest hurricanes, and directly incorporating variables from the Social Vulnerability Index (SVI) into a zone-to-zone (census tract level) evacuation demand model. We find that vehicle availability, residence in group quarters, road density, and English proficiency have significant effects on evacuation demand, shaping both the ability to evacuate from origin tracts and the attractiveness of destination tracts. Travel impedance, measured as distance, also plays a significant role, with evacuees substantially less likely to travel longer distances.
\end{abstract}

\newpage

\section{Introduction}

Hurricanes have long presented challenges to society. These powerful storms disrupt daily life, cause extensive property damage, and pose significant threats to human health and safety. Hurricane Ian, which made landfall in Florida on September 28, 2022, as a category 4 hurricane, is a ruinous example of this. Ian caused \$113 billion in damage and 150 fatalities in Florida alone, making it the costliest and one of the deadliest hurricanes in Florida's history \citep{NationalHurricaneCenter2023}. Despite widespread awareness of the risks, many Floridians chose not to evacuate during hurricanes, even when mandates were in place \citep{Fleischer2022, Olivo2022}. According to a 2024 survey, 33\% of Floridians who chose not to evacuate did so to stay and fix potential damage to their homes, while 29\% believed the storm would change direction, as had happened in previous years \citep{Fleischer2022}. The extensive destruction caused by Hurricane Ian highlights the complex nature of evacuation decisions and the need to analyze evacuation behavior to improve future preparedness.

Understanding hurricane evacuation behavior is critical for effective emergency planning and response. Prior research has largely examined general evacuation behavior, drawing on post-event surveys and interviews as well as administrative records from shelters and hotels to understand whether households evacuate, where they go, and the factors shaping those decisions \citep{Collins2018, Elder2007, Whitehead2000, Cheng2011}. Other studies have emphasized behavioral, spatial, and organizational factors when examining hurricane evacuation behavior. Household- and individual-level research highlights risk perception, caregiving responsibilities, prior experience, and destination type choice \citep{Brodar2020, MesaArango2013}, while spatially oriented studies focus on travel distance, accessibility, and mobility patterns using destination-choice or gravity-based frameworks \citep{Cheng2008, Hong2020, Jiang2021}. Organizational studies further examine evacuation decision-making in institutional settings such as nursing homes, emphasizing preparedness and regulatory constraints rather than household mobility \citep{Peterson2020}.

However, a gap remains in understanding the role of social vulnerability in shaping where evacuees go, and which underlying factors influence where evacuees relocate. This study addresses this gap by disaggregating the socioeconomic variables that comprise the Social Vulnerability Index (SVI) developed by the Centers for Disease Control and Prevention (CDC), and assessing which of these variables significantly influence evacuation flows between census tracts. Through this approach, we model how social vulnerability shapes evacuation destination choice and identify opportunities to better account for these disparities in evacuation planning and response. In response to this gap, we pose the following three research questions:

\begin{enumerate}

  \item What were the dominant spatial patterns of evacuees during Hurricane Ian? Specifically, where did most evacuees go, and how far did they travel from their home?
  \item How did travel impedance and the built environment, particularly distance and road network characteristics, shape evacuation flows and destination choices during Hurricane Ian?
  \item How does social vulnerability influence spatial evacuation patterns and the selection of evacuation destinations?
  
\end{enumerate}

To answer these questions, we analyze large-scale mobile device data to track evacuee movement during Hurricane Ian, offering more precise insights into destination choices than prior studies using geo-tagged tweets and aggregated mobility data \citep{Kumar2018, Li2024}. Traditional methods relying on static datasets, such as census data, surveys, and traffic sensors, struggle to capture dynamic evacuation patterns. Surveys---used in studies such as \citet{Jiang2021}, \citet{Kang2007}, \citet{Lindell2011}, \citet{Solis2010}, \citet{Pham2020}, and \citet{Su2020}---are prone to recall bias \citep{Wilmot2006}, while census data lacks real-time evacuation insights \citep{Cheng2011}. By leveraging mobile device data, we overcome these limitations, providing a more accurate and granular analysis of evacuation behavior.

To model evacuation destination selection, we employ a multiplicative direct-demand model estimated in log-linear form. Direct-demand models relate observed trip flows directly to accessibility and impedance measures between origin--destination (OD) pairs \citep{Talvitie1973}, providing a flexible framework for examining how characteristics at both origins and destinations shape aggregate movement patterns. This approach has been widely applied in transportation research, including station-to-station transit ridership analysis, where OD-based direct-demand models capture the joint influence of origin and destination attributes on observed flows \citep{Choi2012}. More recently, direct-demand formulations have been applied to hurricane evacuation behavior, emphasizing physical and temporal drivers such as travel distance, departure timing, storm surge, and wind speed to explain and predict evacuation patterns \citep{Anyidoho2023, Feng2022}.

This capability helps address the previously identified gap in evacuation modeling by explicitly integrating social vulnerability indicators that are known to influence evacuation behavior \citep{Sun2024}. Although prior studies have shown that incorporating measures of social vulnerability improves the ability to capture evacuation patterns and disparities \citep{Hofflinger2019, Karaye2019, Pence2018}, these factors have not been directly embedded within a direct-demand modeling framework. This study advances the literature by explicitly incorporating social vulnerability variables into a direct-demand model, allowing their influence on evacuation destination choice to be evaluated in a transparent and interpretable manner.

This paper is structured as follows: Section~\ref{sec:lit} reviews relevant literature on evacuation destination choice modeling, the application of gravity models, and the role of social vulnerability in disaster response studies. Section~\ref{sec:methods} describes methodology, including the extraction of mobile device location data and the development of a multiplicative model to analyze evacuation patterns. Section~\ref{sec:data} discusses the case study of Hurricane Ian and the data used. Section~\ref{sec:results} presents the results of the model and provides descriptive statistics on evacuation behavior. Section~\ref{sec:discussion} offers a discussion of the key findings and their implications for evacuation planning and disaster management, as well as clear answers to the three research questions described above. Finally, Section~\ref{sec:conclusion} concludes with a summary of contributions, limitations, and directions for future research.

\section{Literature Review}
\label{sec:lit}

\subsection{Direct demand modeling in evacuation studies}

The direct demand models offer great flexibility to integrate trip generation, trip distribution, and mode split into a single equation. First proposed by \citet{Kraft1967}, demand models are better suited for travel forecasting because they embody a theory of behavior, allowing observed flows to be expressed as multiplicative functions of attributes at the origin, destination, and the connecting travel cost \citep{Talvitie1973}. This advantage creates a more direct linkage between spatial and social characteristics and evacuation flows, giving a much more comprehensive overview of what factors specifically shape evacuation behavior.

Much of the existing hurricane evacuation literature examines evacuation behavior by focusing on specific components of the decision process, such as whether households evacuate, when they depart, or where they go, often in isolation. For example, studies emphasizing economic considerations have shown that income, evacuation costs, and expected storm damage significantly influence evacuation participation and accommodation choices \citep{Whitehead2000}. Other research has focused on household characteristics, including household size, vehicle availability, length of residence, and prior hurricane experience, demonstrating how mobility constraints and social ties shape evacuation decisions \citep{Smith2009, Carnegie2010, Deka2012, MesaArango2013}.

Other studies have concentrated on departure timing, identifying proximity to hazards, evacuation orders, anticipated congestion, and household mobility resources as key determinants of when evacuees leave \citep{Fu2006, Gudishala2012}. Other destination-focused studies have examined spatial patterns of evacuation destinations using aggregated zones or distance-based measures, highlighting the roles of accessibility and roadway structure but often without explicitly modeling destination attributes themselves \citep{Radwan2005}.

While these studies collectively identify a wide range of factors influencing evacuation behavior that are immensely insightful, they typically model these decisions independently, treating departure time, accommodation choice, destination choice, or travel distance as separate outcomes. As a result, they offer limited insight into how origin-side vulnerabilities, destination-side characteristics, and travel impedance jointly shape observed evacuation flows. Direct-demand modeling addresses this gap by explicitly connecting these elements within a single origin--destination framework, allowing for a more integrated assessment of how social vulnerability, built-environment features, and distance interact to influence evacuation destination choice and flow intensity.

Others have already begun using direct demand modeling for this purpose, but often emphasize physical hazard conditions and timing of evacuation behavior. \citet{Feng2022} applied a direct-demand framework to evacuation flows during hurricanes by incorporating travel distance, departure timing, storm surge exposure, and wind intensity as predictor variables. Their model highlighted the dominant role of hazard intensity and travel impedance in shaping evacuation volumes, particularly under rapidly evolving storm conditions. Similarly, \citet{Anyidoho2023} used a flow-based modeling approach to predict evacuation demand by integrating meteorological variables, evacuation order timing, and roadway characteristics, demonstrating that OD-level demand models can effectively capture aggregate evacuation responses to storm dynamics.

These studies, while providing important advancements in direct demand modeling, focus on physical exposure, storm timing, and network-related constraints; they neglect the social and demographic characteristics that could be influencing these evacuation flows. Building on these prior efforts, this study extends direct demand modeling by explicitly incorporating formal social vulnerability indicators and built-environment measures at both the origin and destination. By doing so, the model moves beyond purely physical or temporal drivers and provides a more comprehensive representation of evacuation destination choice.

\subsection{Social vulnerability and evacuation behavior}

To systematically incorporate social vulnerability into evacuation modeling, we choose to use the Social Vulnerability Index (SVI), a well-established metric developed by the Centers for Disease Control and Prevention (CDC). The SVI quantifies community vulnerability at the census tract level, making it highly compatible with spatial models of evacuation behavior \citep{CDC2024}. Unlike other measures, the SVI consolidates multiple socioeconomic and demographic factors into a single metric, providing a comprehensive representation of the structural conditions that influence households' capacity to prepare for, respond to, and recover from hazards.

While not explicitly designed for evacuation modeling, the SVI aligns with key factors affecting evacuation behavior, such as household composition, transportation access, and disability status. Prior studies confirm that social vulnerabilities shape both evacuation destinations and means of transportation \citep{Hofflinger2019, Karaye2019, Pence2018, Sun2024}. The SVI consists of four themes: 1) Socioeconomic Status; 2) Household Characteristics; 3) Minority Status \& Language; and 4) Housing Type \& Transportation. In total, 32 variables comprise the SVI, and all will be considered in this study at both the origin, $i$, and destination, $j$, tracts ($n = 64$).

Socially vulnerable populations often face compounded barriers during evacuations, including limited access to private vehicles, financial constraints, mobility limitations, and challenges in receiving or acting upon emergency information \citep{Flanagan2011}. These constraints can delay evacuation, restrict destination options, or force households to rely on nearby or socially connected locations rather than formally designated shelters. Consequently, social vulnerability not only influences evacuation participation but also shapes evacuation pathways and outcomes. Incorporating social vulnerability into evacuation modeling is therefore essential for capturing inequities in mobility and safety and for developing policies that better support at-risk populations during extreme events.

\section{Methods}
\label{sec:methods}

Based on large-scale mobile device location data consisting of millions of anonymized GPS points, we implement a multi-step methodological framework to analyze evacuation behavior during Hurricane Ian. First, we apply a home detection algorithm developed by \citet{Zhang2024a} to identify likely residents and estimate their home locations prior to the storm, which serve as the origins for evacuation trips. This algorithm uses temporal and spatial patterns of device presence to infer residential locations with high accuracy. Next, we develop a destination inference algorithm to filter and analyze GPS data collected during the evacuation period. This step involves identifying significant location clusters visited after departure, applying temporal thresholds to confirm overnight stays, and removing transient or pass-through locations. Finally, we construct and calibrate the multiplicative model to fit the observed origin--destination (OD) flows.

\subsection{The direct demand model}

Direct-demand models have been widely used in transportation research to evaluate how observed trip flows respond to characteristics of origins, destinations, and travel impedance. These models are commonly estimated using ordinary least squares (OLS) regression, which can accommodate both continuous and categorical variables, is flexible and computationally efficient, and yields coefficients that are straightforward to interpret \citep{Talvitie1973, Kuby2004, Hannay2007, Manaugh2010, Yan2020}. In this study, we adopt a direct-demand formulation to model evacuation flows between census tracts during Hurricane Ian. The general functional form is given by:

\begin{equation}
T_{ij} = f\left(L_{ij},\, \mathbf{SED}_i,\, \mathbf{BE}_i,\, \mathbf{SED}_j,\, \mathbf{BE}_j\right)
\label{eq:general}
\end{equation}

\noindent where $T_{ij}$ denotes the observed evacuation flow count from origin tract $i$ to destination tract $j$; $L_{ij}$ represents the set of travel-impedance variables between $i$ and $j$ (e.g., distance between tract centroids); $\mathbf{SED}_i$ and $\mathbf{SED}_j$ denote vectors of socioeconomic and demographic characteristics for the origin and destination tracts, respectively; and $\mathbf{BE}_i$ and $\mathbf{BE}_j$ present built-environment characteristics of the origin and destination tracts. This formulation allows evacuation flows to be modeled directly as a function of origin-side constraints, destination-side attributes, the built environment, and travel impedance without relying on balancing constraints or composite accessibility indices.

\subsection{The multiplicative model}

The earliest forms of direct-demand models were multiplicative \citep{Ortuzar2001, Talvitie1973}. A key advantage of the multiplicative formulation is its ability to model origin-side and destination-side effects separately, allowing the impacts of conditions at evacuation origins to be distinguished from those at evacuation destinations \citep{Choi2012}. Prior work has shown that multiplicative direct-demand models can outperform alternative count-based specifications, such as Poisson regression, in explaining station-to-station travel flows \citep{Choi2012}. Motivated by these advantages, we adopt a multiplicative direct-demand model to analyze evacuation flows during Hurricane Ian. The model is defined as:

\begin{equation}
T_{ij} = \phi \cdot \prod_{p=1}^{P} X_{ip}^{\alpha_p} \cdot \prod_{p=1}^{P} X_{jp}^{\beta_p} \cdot \prod_{q=1}^{Q} Z_{ijq}^{\gamma_q}
\label{eq:multiplicative}
\end{equation}

\noindent where $X_{ip} \in \mathbf{SED}_i \cup \mathbf{BE}_i \cup \mathbf{TS}_i$ is the $p$th feature regarding origin zone $i$, $X_{jp} \in \mathbf{SED}_j \cup \mathbf{BE}_j \cup \mathbf{TS}_j$ is the $p$th feature regarding destination zone $j$, $Z_{ijq} \in L_{ij}$ is the $q$th feature regarding the travel impedance from $i$ to $j$, $\phi$ is the scale parameter, $\alpha_p$, $\beta_p$, $\gamma_q$ are the parameters to be estimated, $P$ is the total number of variables that include socioeconomic and demographic and built-environment variables, and $Q$ is the total number of travel-impedance variables. Then, by taking the natural logarithm on both sides of Eq.~(\ref{eq:multiplicative}), we transform it into linear form:

\begin{equation}
\ln T_{ij} = \ln\phi + \sum_{p=1}^{P} \alpha_p \ln X_{ip} + \sum_{p=1}^{P} \beta_p \ln X_{jp} + \sum_{q=1}^{Q} \gamma_q \ln Z_{ijq}
\label{eq:loglinear}
\end{equation}

To ensure results are interpretable on the original flow scale, all predictive metrics were computed after back-transforming predicted values from the log-linear model form, $\ln \hat{T}_{ij}$, to the original evacuee-count domain, $\hat{T}_{ij} = e^{\ln\hat{T}_{ij}}$.

\subsection{Model estimation and validation}

We then estimated a multiplicative direct-demand model in log-linear form using ordinary least squares (OLS), with evacuation flow counts between each origin--destination pair ($T_{ij}$) as the response variable and the predictors screened using the variance inflation factor (VIF) as covariates. Model fitting and evaluation were implemented in Python (\texttt{StatsModels} and \texttt{scikit-learn}). Model inference was based on the full-sample OLS fit, and we identified influential predictors using statistical significance ($p < 0.05$) for subsequent interpretation and discussion.

To assess predictive performance, we apply a 10-fold cross-validation, in which the dataset was randomly partitioned into 10 disjoint subsets. Each fold was held out in turn while the model was trained on the remaining nine folds, and predictions of evacuation flow counts were generated for the holdout fold. This process was repeated for all 10 folds, and performance metrics were averaged to obtain mean in-sample and out-of-sample estimates. We report the following validation metrics: Root Mean Squared Error (RMSE), Mean Absolute Error (MAE), and $R^2$.

The \textbf{Root Mean Squared Error (RMSE)} is derived from the Mean Squared Error (MSE), which quantifies the average of the squared differences between the predicted and actual values:

\begin{equation}
\mathrm{RMSE} = \sqrt{\frac{1}{N}\sum_{(i,j)}\left(\hat{T}_{ij} - T_{ij}\right)^2}
\end{equation}

\noindent where $\hat{T}_{ij}$ denotes the predicted evacuation flow count for each origin-destination pair $(i,j)$, $T_{ij}$ is the corresponding observed evacuation flow count, and $N$ is the total number of OD pairs. RMSE reflects the typical magnitude of prediction error in units of evacuees.

The \textbf{Mean Absolute Error (MAE)} measures the average absolute differences between the predicted and actual values:

\begin{equation}
\mathrm{MAE} = \frac{1}{N}\sum_{(i,j)}\left|\hat{T}_{ij} - T_{ij}\right|
\end{equation}

\begin{figure}[H]
  \centering
  \includegraphics[width=\textwidth]{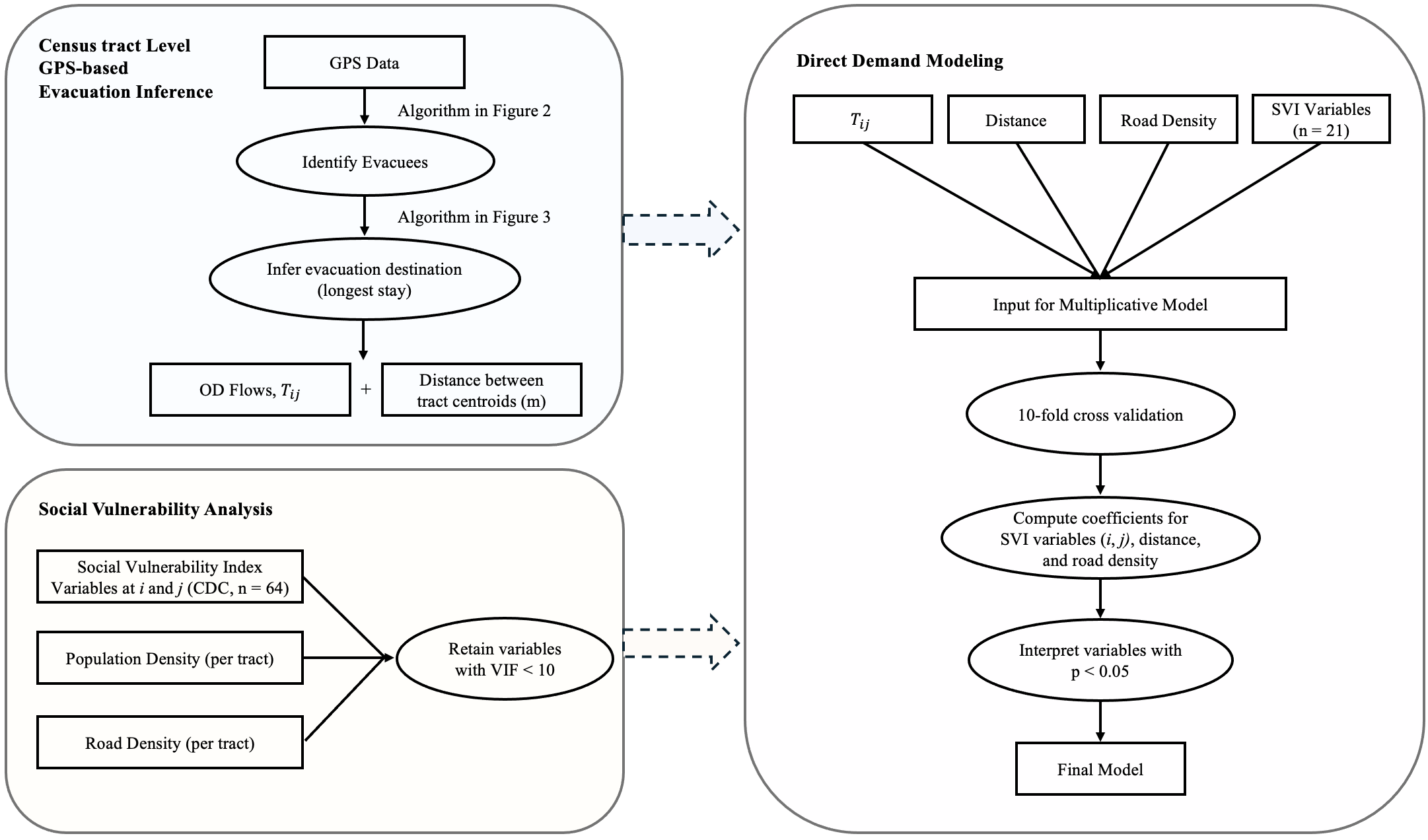}
  \caption{Method overview.}
  \label{fig:method}
\end{figure}

Finally, the \textbf{$R$-squared} ($R^2$) value assesses the proportion of the variance in the dependent variable that is explained by the model:

\begin{equation}
R^2 = 1 - \frac{\sum_{(i,j)}\left(\hat{T}_{ij} - T_{ij}\right)^2}{\sum_{(i,j)}\left(T_{ij} - \bar{T}\right)^2}
\end{equation}

\noindent where $\bar{T}$ is the mean observed evacuation flow across all origin--destination pairs. Together, these metrics provide a comprehensive evaluation of the model's ability to capture both the magnitude and spatial variation of evacuation flows. The overall study framework is shown in Fig.~\ref{fig:method}.

\section{Case Study and Data}
\label{sec:data}

Our case study focuses on Hurricane Ian, which ranks as the third costliest and one of the deadliest hurricanes in United States history \citep{NOAA2022}. This devastating Category 5 hurricane landed in Lee County, which accounted for nearly half of the deaths and reportedly issued evacuation orders less than 36 hours before Ian's landfall \citep{Fleischer2022}. This has raised numerous concerns about the local government's response time, suggesting that delays in information release contributed to the high fatality rate \citep{Verma2024}. Therefore, in this paper, we focus on the evacuation activities of residents in Lee County, Florida.

The mobile device location dataset used in this study covers the whole state of Florida, spanning from September 23, 2022, to October 15, 2022. The fields in the raw data contain unique device identifiers, latitude, longitude, timestamps, GPS errors, etc. To ensure data quality and exclude users with low-frequency data, we remove duplicate points and any users with fewer than 150 GPS points. We retain only medium-high to high accuracy points as defined by Gravy Analytics, where GPS errors do not exceed 50 m \citep{Mryan2022, Liu2025}. After data cleaning, we included 2,286,251,350 GPS records generated by 9,242,974 unique devices. All device identifiers were further processed with de-identification techniques and were ensured to contain no sensitive information.

\subsection{Home detection}

We first distinguish Florida residents from transient devices and identify each resident's home location prior to Hurricane Ian. We adopt the home detection algorithm developed by \citet{Zhang2024a}, which has demonstrated high performance in similar applications \citep{Verma2024}. Assuming evacuees begin their trips from home, we aggregated each user's GPS points into 20-by-20-meter grids and calculated stay durations per cell. A home location was assigned to the cell where users spent the most time during nights and weekends, with a minimum of five nights required \citep{Liu2025}. If no night data were available, we assigned home based on a minimum of six hours spent in a cell during weekends. We then define "active days" as days with at least 10 GPS points. Devices with fewer than 15 active days during the pre-Ian period are discarded to avoid spurious home assignments. The remaining devices are classified as Florida residents, and their assigned home locations (grid cell centroids) are used as the starting points for potential evacuation trips. All home locations are inferred exclusively from pre--Hurricane Ian data to distinguish evacuation-related relocations from routine mobility patterns, ensuring that observed movements reflect hurricane-induced displacement rather than normal daily travel \citep{Li2024}.

\subsection{Evacuation destination inference}

\begin{figure}[H]
  \centering
  \includegraphics[width=0.8\textwidth]{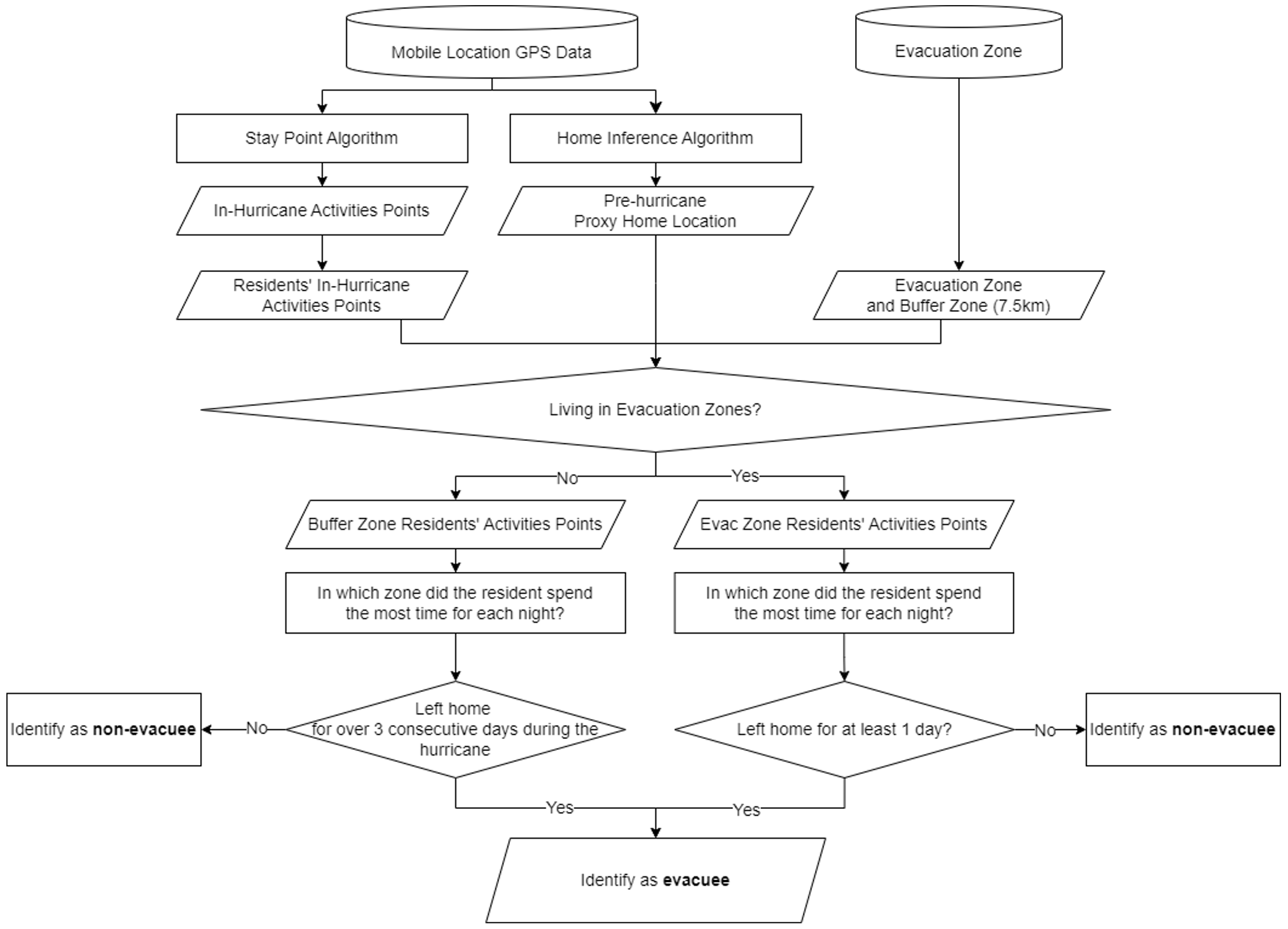}
  \caption{The chart flow of the evacuation behavior inference algorithm.}
  \label{fig:evac_inference}
\end{figure}

With the home location and the movement trajectory of residents in Florida, we inferred the destination of evacuees with two algorithms. The first is the evacuation behavior inference algorithm as shown in Fig.~\ref{fig:evac_inference}, which identifies the evacuees out of the residents based on their moving trajectory and the evacuation order status in their corresponding residential area \citep{Liu2025}. We first collected the evacuation zone and order information and generated the designated zones for evacuation. Meanwhile, due to the existence of shadow evacuees \citep{Liu2025, Lovreglio2020} who evacuated despite living out of the designated zones, we also generated a 7.5 km buffer zone outside the designated evacuation zone. Based on whether a resident lives in or out of the designated evacuation zone, we then filtered the activities (inferred by algorithms proposed by \citet{Zhang2024b}) throughout the whole trajectory and detected which zone the resident spent the most time in each night. We also removed the residents who did not generate any data points for more than half of the night during the hurricane. If a resident inside the evacuation zone left home for one night or a resident inside the buffer zone left home for three consecutive nights, then we identify the user as an evacuee.

\begin{figure}[H]
  \centering
  \includegraphics[width=0.8\textwidth]{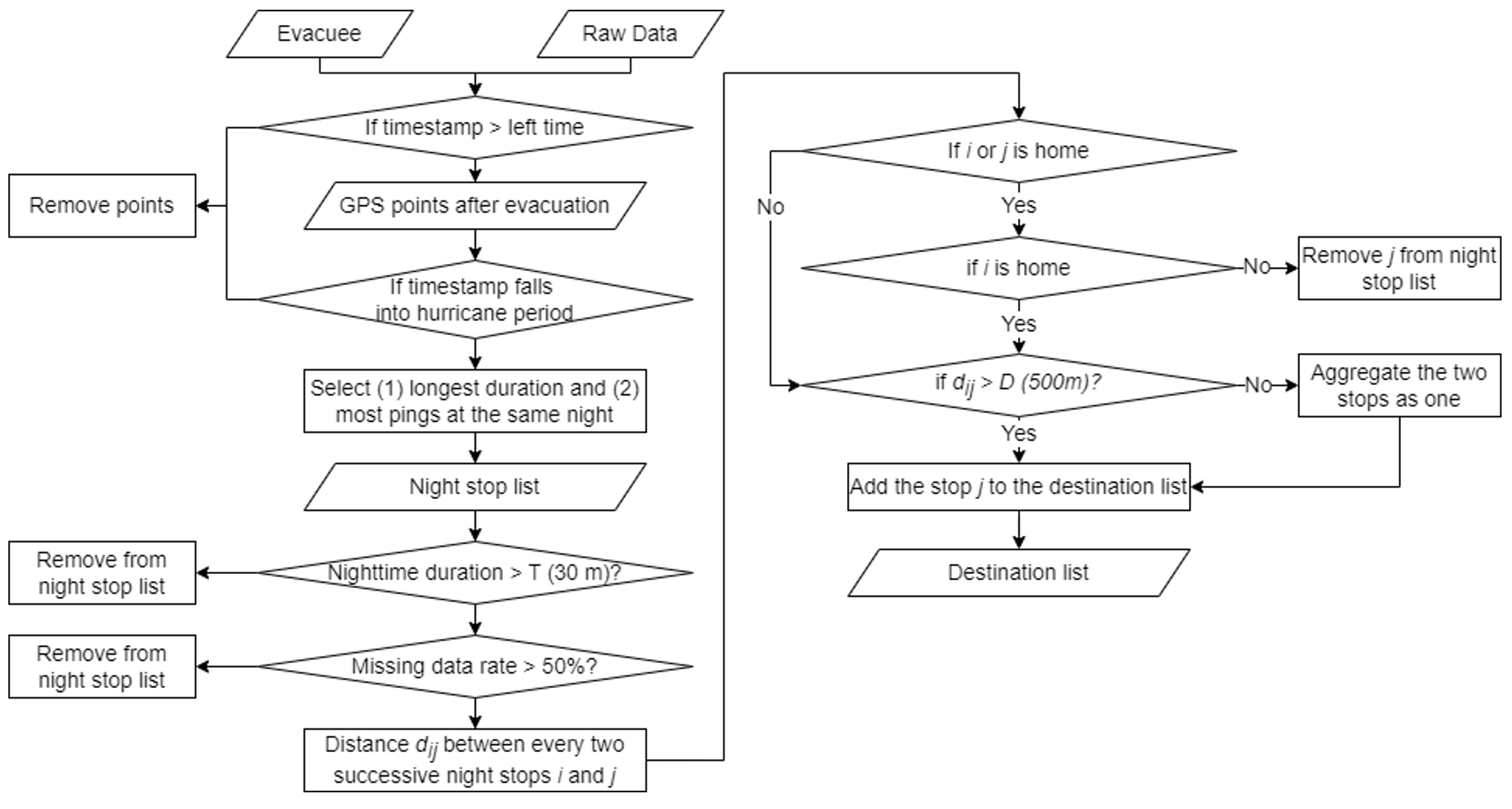}
  \caption{The chart flow of the evacuation destination inference algorithm.}
  \label{fig:dest_inference}
\end{figure}

The second algorithm is the evacuation destination inference algorithm, which identifies the destination of evacuees after they leave their homes. Figure~\ref{fig:dest_inference} presents a flowchart of the procedure. Based on the list of evacuees, we first sorted their GPS points recorded after departure from home and within the hurricane period (8~PM September~22 to 7~AM October~1,~2022). We then adopted a grid-based approach to infer the stops of each individual. Specifically, we overlaid each user’s GPS points onto a 20~m $\times$ 20~m grid and identified the grid cell with the highest number of pings for each night (using the longest duration as a tiebreaker when multiple cells had the same number of pings). Next, we calculated the distance between consecutive stops. If a stop exceeded a minimum distance threshold from either the user’s home location or the previous stop, it was retained and added to the stop list for that evacuee.

With all destinations of each user, we first identified the census tracts corresponding to each individual user’s origin and destination coordinates. We then selected the destination census tract where each user stayed for the longest number of consecutive nights as their evacuation destination. This approach ensures that we captured the primary evacuation destination rather than transient stops, providing a more accurate reflection of the final evacuation decisions \citep{Lovreglio2020}.

To further validate the accuracy of the inferred home and destination locations, we cross-matched these points against statewide land-use classifications. Land-use polygons were obtained from the Florida Department of Environmental Protection (FDEP) Statewide Land Use / Land Cover dataset, which compiles data from five regional land-use inventories. The dataset contains approximately 1.43~million polygons, is publicly available through the FDEP Open Data Portal, and is periodically updated (most recently on July~18,~2025). We performed spatial joins between each user’s home and destination coordinates and the Level-1 land-use classes (\texttt{LEVEL1\_LAN}) to determine whether these locations fall within residential or non-residential parcels. 

Within this classification system, the \texttt{LEVEL1\_LAN = 1000} category ("Urban and Built-Up") encompasses a range of Level-2 subcategories, including low-, medium-, and high-density residential areas, as well as commercial, institutional, service, and recreational uses. This broad classification provides a useful basis for interpreting evacuation destinations, including movements to residential neighborhoods, commercial lodging, or potential shelter locations. In this study, parcels coded as \texttt{LEVEL1\_LAN = 1000} were classified as residential, while all other codes were categorized as non-residential or unmatched.

Results indicate that approximately 83\% of inferred home locations and 82\% of inferred destination locations fall within residential land-use parcels. This strong alignment between GPS-derived locations and expected residential land-use patterns supports the validity of both the home-detection procedure and the destination inference algorithm.

We note a temporal mismatch between the datasets used in this validation: the GPS mobility data reflect evacuation behavior during Hurricane Ian in 2022, whereas the land-use dataset was released in 2025. Although broad residential land-use patterns are generally stable over short time horizons, localized development or reclassification may have occurred. Therefore, this validation should be interpreted as an approximate consistency check rather than a parcel-level ground truth.

\begin{figure}[H]
  \centering
  \includegraphics[width=\textwidth]{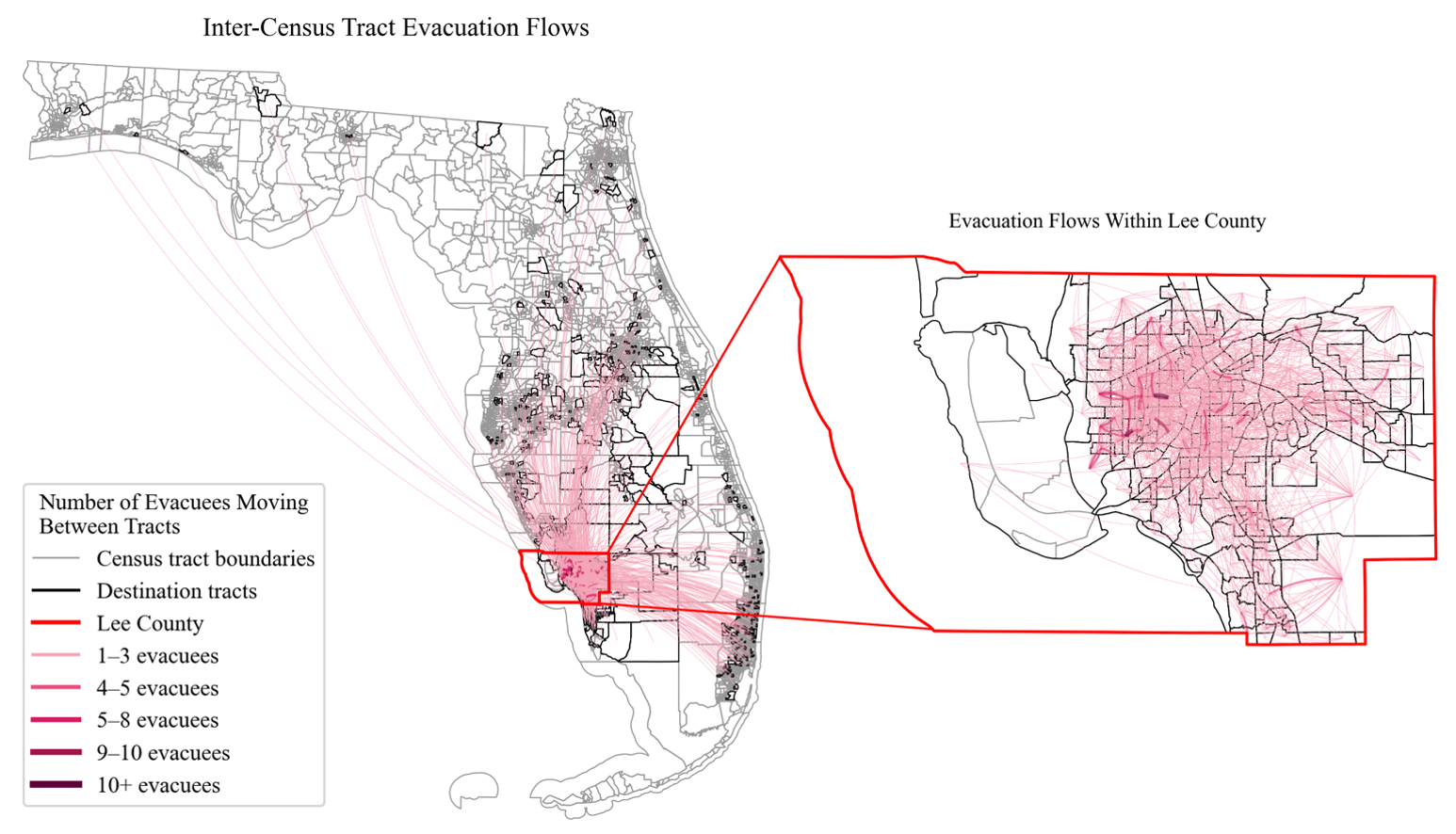}
  \caption{\textit{Inter- and intra-county origin--destination (OD) flow maps of evacuee travel during Hurricane Ian.} The left panel illustrates inter-county movements across Florida, with lines representing trips between census tracts, most of which were to nearby areas. The right panel zooms in on Lee County, highlighting dense intra-county flows.}
  \label{fig:OD_flows}
\end{figure}

Using the identified home and destination census tracts for each evacuee, we constructed a tract-level origin--destination (OD) matrix from the GPS-based mobility data. Text-based OD records linking home tracts to destination tracts with associated evacuee counts (e.g., "12071000303 $\rightarrow$ 12015010100 = 1") were aggregated such that each row represents a unique tract pair and its observed flow. These OD flows serve as the response variable in the model calibration (shown in Figure~\ref{fig:OD_flows}).

\subsection{Social vulnerability data}

In addition to mobile location data, we incorporate the full Social Vulnerability Index (SVI) at the census tract level by including all 32 component variables provided by the Centers for Disease Control and Prevention (CDC) at the origin and destination tract, totaling 64 variables considered in this study. These variables capture four dimensions of vulnerability: socioeconomic status, household composition and disability, minority status and language, and housing type and transportation. Rather than relying on composite SVI theme scores or a limited subset of indicators, this approach allows us to explicitly examine which individual SVI components, measured separately for origin and destination tracts, are associated with evacuation flows within the multiplicative modeling framework.

In addition to these SVI measures, we constructed three OD "connector" and built-environment predictors: (i) centroid-to-centroid distance between origin and destination tracts (meters), (ii) population density for each tract (people per km$^2$) on both the origin and destination side, and (iii) road density for each tract (km of roads per km$^2$) on both the origin and destination side. All variables were joined to the OD flow table using the corresponding origin and destination tract identifiers. Table~\ref{tab:variables} provides variable descriptions for all predictors initially considered in the analysis.

\begin{table}[H]
\centering
\caption{Variable Description}
\label{tab:variables}

\small
\setlength{\tabcolsep}{4pt}
\renewcommand{\arraystretch}{0.95}

\begin{tabular}{p{3.2cm} p{8.8cm}}
\toprule
\textbf{Variable} & \textbf{Description} \\
\midrule

\multicolumn{2}{c}{\textit{\textbf{Theme 1: Socioeconomic Status}}} \\
E\_POV150 & Persons below 150\% of poverty level \\
E\_UNEMP & Unemployed civilians age 16+ \\
E\_HBURD & Housing cost-burdened units (income $<\$75$k) \\
E\_NOHSDP & Persons age 25+ with no high school diploma \\
E\_UNINSUR & Uninsured civilian noninstitutionalized population \\
EP\_POV150 & Percent below 150\% poverty \\
EP\_UNEMP & Unemployment rate (\% age 16+ labor force) \\
EP\_HBURD & Percent cost-burdened ($<\$75$k; $\geq$30\% income) \\
EP\_NOHSDP & Percent with no high school diploma \\
EP\_UNINSUR & Percent uninsured population \\

\midrule
\multicolumn{2}{c}{\textit{\textbf{Theme 2: Household Composition \& Disability}}} \\
E\_AGE65 & Persons age 65+ \\
E\_AGE17 & Persons age 17 and younger \\
E\_DISABL & Population with a disability \\
E\_SNGPNT & Single-parent households with children $<18$ \\
E\_LIMENG & Persons age 5+ speaking English less than well \\
EP\_AGE65 & Percent age 65+ \\
EP\_AGE17 & Percent age 17 and younger \\
EP\_DISABL & Percent with a disability \\
EP\_SNGPNT & Percent single-parent households \\
EP\_LIMENG & Percent limited English proficiency \\

\midrule
\multicolumn{2}{c}{\textit{\textbf{Theme 3: Minority Status \& Language}}} \\
E\_MINRTY & Minority population (non-Hispanic minorities + Hispanics) \\
EP\_MINRTY & Percent minority population \\

\midrule
\multicolumn{2}{c}{\textit{\textbf{Theme 4: Housing Type \& Transportation}}} \\
E\_MUNIT & Housing units in structures with 10+ units \\
E\_MOBILE & Mobile homes \\
E\_CROWD & Occupied units with more people than rooms \\
E\_NOVEH & Households with no vehicle available \\
E\_GROUPQ & Persons in group quarters \\
EP\_MUNIT & Percent multi-unit structures (10+) \\
EP\_MOBILE & Percent mobile homes \\
EP\_CROWD & Percent crowded households \\
EP\_NOVEH & Percent households without vehicles \\
EP\_GROUPQ & Percent in group quarters \\

\midrule
\multicolumn{2}{c}{\textit{\textbf{Built Environment}}} \\
PDENSITY\_O & Population density (origin; persons/km$^2$) \\
PDENSITY\_D & Population density (destination; persons/km$^2$) \\
RDENSITY\_O & Road density (origin; km/km$^2$) \\
RDENSITY\_D & Road density (destination; km/km$^2$) \\

\midrule
\multicolumn{2}{c}{\textit{\textbf{Travel Impedance}}} \\
DISTANCE & Distance between tract centroids (meters) \\

\midrule
\multicolumn{2}{c}{\textit{\textbf{OD Flow}}} \\
$T_{ij}$ & Flow from tract $i$ (origin) to tract $j$ (destination) \\

\bottomrule
\end{tabular}
\end{table}

To reduce multicollinearity prior to model estimation, we computed variance inflation factors (VIFs) on the full predictor set and iteratively removed any variable with VIF $>$ 10. We applied a $\log(x+1)$ transformation to any predictor containing zeros before computing VIFs and estimating models. After VIF-based screening, the final predictor set comprises 21 variables, including OD distance, road density, and retained SVI components from all themes except Theme 3. Table~\ref{tab:descriptive} displays the descriptive statistics for these variables.

\begin{table}[H]
\centering
\caption{Descriptive Statistics of Variables Used for Modeling}
\label{tab:descriptive}

\small
\setlength{\tabcolsep}{5pt}
\renewcommand{\arraystretch}{0.95}

\begin{tabular}{p{3.2cm} p{2.2cm} p{2cm} p{2cm} p{2cm} p{2cm}}
\toprule
\textbf{Variable} & \textbf{Side} & \textbf{Mean} & \textbf{SD} & \textbf{Min} & \textbf{Max} \\
\midrule

\multicolumn{6}{c}{\textit{\textbf{Theme 1: Socioeconomic Status}}} \\
EP\_UNEMP & Origin & 4.487 & 3.635 & 0 & 17.4 \\
EP\_UNEMP & Destination & 4.472 & 3.614 & 0 & 25.1 \\

\midrule
\multicolumn{6}{c}{\textit{\textbf{Theme 2: Household Composition \& Disability}}} \\
EP\_SNGPNT & Origin & 4.696 & 5.211 & 0 & 37 \\
EP\_SNGPNT & Destination & 4.927 & 5.299 & 0 & 37 \\
EP\_LIMENG & Origin & 5.309 & 6.044 & 0 & 33.4 \\
EP\_LIMENG & Destination & 5.329 & 6.294 & 0 & 46.1 \\

\midrule
\multicolumn{6}{c}{\textit{\textbf{Theme 4: Housing Type \& Transportation}}} \\
E\_GROUPQ & Origin & 120.901 & 536.545 & 0 & 3634 \\
E\_GROUPQ & Destination & 80.425 & 411.918 & 0 & 13835 \\
EP\_MUNIT & Origin & 15.183 & 19.032 & 0 & 89.3 \\
EP\_MUNIT & Destination & 15.695 & 19.984 & 0 & 99.2 \\
EP\_MOBILE & Origin & 8.560 & 18.673 & 0 & 98.5 \\
EP\_MOBILE & Destination & 8.120 & 17.115 & 0 & 98.5 \\
EP\_CROWD & Origin & 2.238 & 3.100 & 0 & 13.2 \\
EP\_CROWD & Destination & 2.512 & 3.472 & 0 & 27.1 \\
EP\_NOVEH & Destination & 5.231 & 6.156 & 0 & 38 \\
EP\_GROUPQ & Origin & 3.216 & 14.266 & 0 & 95.8 \\
EP\_GROUPQ & Destination & 1.912 & 7.898 & 0 & 100 \\

\midrule
\multicolumn{6}{c}{\textit{\textbf{Built Environment}}} \\
RDENSITY\_O & Origin & 1.371 & 1.892 & 0 & 10.254 \\
RDENSITY\_D & Destination & 1.467 & 2.035 & 0 & 22.312 \\

\midrule
\multicolumn{6}{c}{\textit{\textbf{Travel Impedance}}} \\
DISTANCE & --- & 36593.21 & 71975.83 & 0 & 677358.18 \\

\midrule
\multicolumn{6}{c}{\textit{\textbf{OD Flow}}} \\
$T_{ij}$ & --- & 2.972 & 8.186 & 1 & 133 \\

\bottomrule
\end{tabular}
\end{table}

\section{Results}
\label{sec:results}

\subsection{Descriptive analysis}

\begin{figure}[H]
  \centering
  \includegraphics[width=0.9\textwidth]{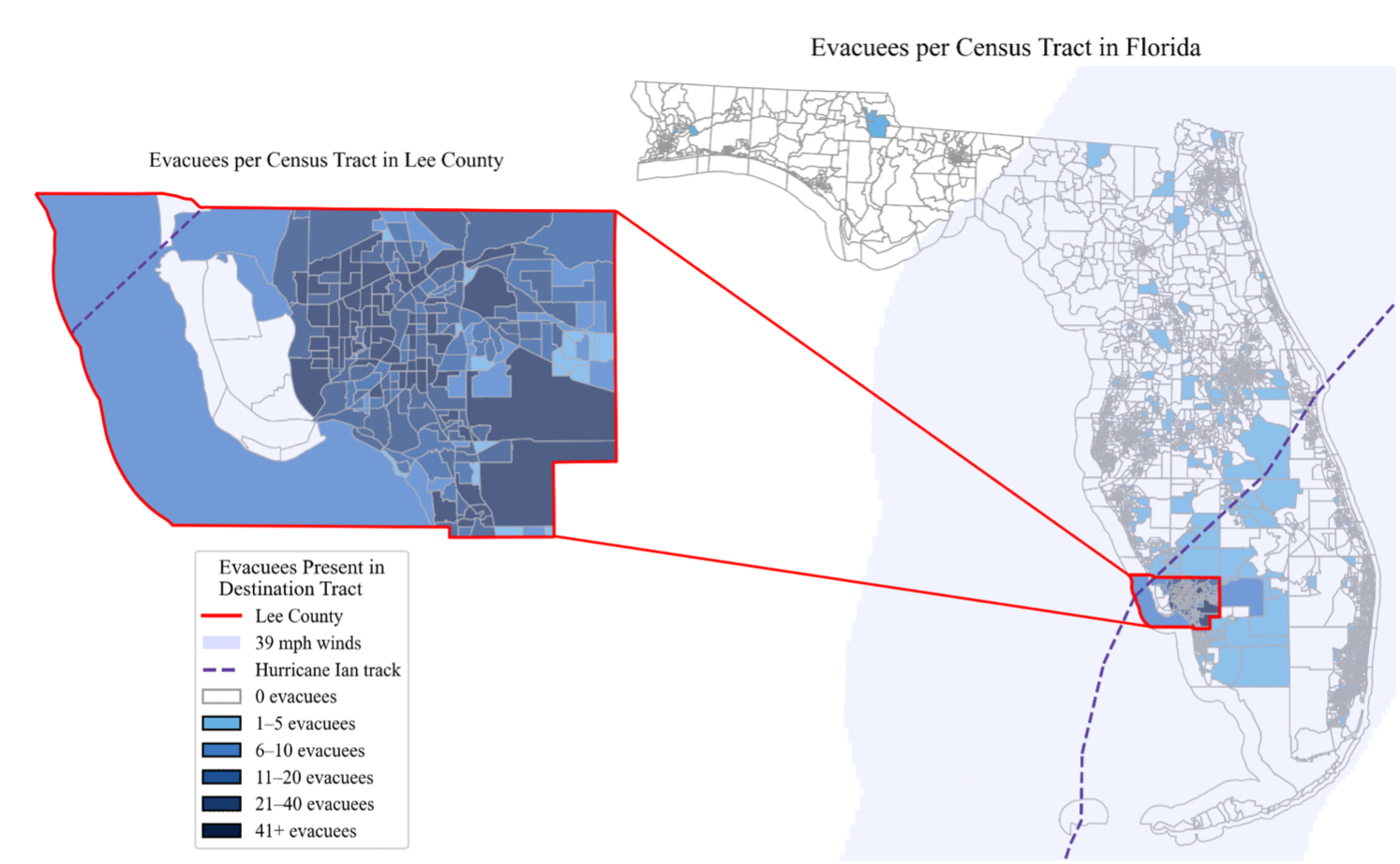}
  \caption{Spatial distribution of evacuees.}
  \label{fig:spatial}
\end{figure}

Our algorithms identified destination choices for 7,980 evacuees from Lee County, Florida, who relocated across 657 census tracts statewide, a dataset substantially larger than prior survey-based studies \citep{Huang2017, Lindell2012}. As shown in Fig.~\ref{fig:spatial}, 92.7\% of evacuees remained within Lee County, clustering around urban areas such as Cape Coral, Fort Myers, and Bonita Springs. Only 7.3\% evacuated outside the county, reflecting a strong preference to stay close to home. Among those who did relocate outside Lee County, Collier County received the largest share---1.69\% of all evacuees. Collier lies roughly 50–60 km south of Lee, making it the nearest large neighboring county with substantial housing stock and available inland shelter options. The main pattern to note is that evacuees overwhelmingly selected destinations that were close and easily accessible (\ref{fig:spatial}).

\subsection{SVI Variables in our model}

We estimated a multiplicative direct-demand model in log-linear form using evacuee counts between each tract pair as the re- sponse variable. Predictors were drawn from the CDC SVI component measures, the built environment, and distance between census tract centroids (m), with a selective transformation applied to each predictor (natural log for strictly positive variables; when a predictor contained any zero values). The final model was estimated on 2,665 OD observations with 21 predictors (excluding the intercept), and model performance was assessed using 10-fold cross validation on the original evacuee-count scale after back- transforming predictions.

All 21 predictors are reported in Table~\ref{tab:model_results}, along with their estimated coefficients, p-values, and levels of statistical significance. Table~\ref{tab:model_results} also summarizes model performance, including in-sample and out-of-sample mean absolute error (MAE), root mean squared error (RMSE), and the coefficient of determination ($R^2$).

\begin{table}[H]
\centering
\caption{Predictor Variable Coefficients, Statistical Significance, and Model Performance}
\label{tab:model_results}

\small
\setlength{\tabcolsep}{5pt}
\renewcommand{\arraystretch}{0.95}

\begin{tabular}{p{9cm} p{2.5cm} p{2.5cm}}
\toprule
\textbf{Variable} & \textbf{Coefficient} & \textbf{p-value} \\
\midrule

\multicolumn{3}{c}{\textit{\textbf{Origin Variables}}} \\
Percent of occupied units with more people than rooms & -0.018764 & 0.2064 \\
Percent of persons in group quarters & -0.007533 & 0.7448 \\
Percent of persons age 5+ who speak English less than well & 0.025285 & 0.05696 \\
Percent of housing units that are mobile homes & 0.007067 & 0.3813 \\
Percent of housing units in structures with 10+ units & 0.001173 & 0.8874 \\
Percent of households with no vehicle available & -0.029619 & 0.01727* \\
Percent of households that are single-parent with children under 18 & -0.013025 & 0.3146 \\
Unemployment rate (age 16+ labor force) & 0.017809 & 0.1974 \\
Persons in group quarters & 0.026202 & 0.00825** \\
Road density (origin; km/km$^2$) & -0.043353 & 0.01844* \\

\midrule
\multicolumn{3}{c}{\textit{\textbf{Destination Variables}}} \\
Percent of occupied units with more people than rooms & 0.014398 & 0.3221 \\
Percent of persons in group quarters & -0.044422 & 0.08354 \\
Percent of persons age 5+ who speak English less than well & 0.029280 & 0.02306* \\
Percent of housing units that are mobile homes & 0.009983 & 0.1914 \\
Percent of housing units in structures with 10+ units & -0.002436 & 0.7666 \\
Percent of households with no vehicle available & -0.015317 & 0.2291 \\
Percent of households that are single-parent with children under 18 & 0.002084 & 0.8705 \\
Unemployment rate (age 16+ labor force) & 0.015216 & 0.2877 \\
Persons in group quarters & 0.023048 & 0.01957* \\
Road density (destination; km/km$^2$) & 0.036927 & 0.0283* \\

\midrule
\multicolumn{3}{c}{\textit{\textbf{Distance}}} \\
Distance between tract centroids (meters) & -0.241234 & $<0.001^{***}$ \\

\midrule
\multicolumn{3}{c}{\textit{\textbf{Model Performance Metrics}}} \\
\multicolumn{3}{c}{In-sample: MAE = 1.573 \quad RMSE = 6.584 \quad $R^2$ = 0.357} \\
\multicolumn{3}{c}{Out-of-sample: MAE = 1.589 \quad RMSE = 6.617 \quad $R^2$ = 0.351} \\

\bottomrule
\end{tabular}

\vspace{0.5em}
\footnotesize
\textit{Note:} * $p < 0.05$, ** $p < 0.01$, *** $p < 0.001$.
\end{table}

Across the full model, origin--destination (OD) distance exhibited a strong negative association with evacuation flows ($\beta = -0.241234$), indicating that longer centroid-to-centroid distances between origin and destination tracts are associated with fewer observed evacuee movements. This pattern is consistent with the spatial concentration of evacuation activity observed in the previous section. 

Beyond distance, several Social Vulnerability Index (SVI) components and built-environment variables exhibited distinct and asymmetric relationships depending on whether they characterize the origin or the destination tract, underscoring the importance of accounting for social vulnerability in evacuation behavior.

On the origin side, the share of households with no vehicle available was negatively associated with evacuation flows ($\beta = -0.029619$), suggesting reduced evacuation capacity in tracts with limited household mobility resources. In contrast, the number of persons living in group quarters at the origin was positively associated with evacuee flows ($\beta = 0.026202$), consistent with elevated evacuation demand among populations residing in institutional or collective living settings. Origin-side road density also exhibited a negative association ($\beta = -0.043353$), indicating that tracts with denser road networks—often corresponding to urban cores—were associated with lower modeled outflows.

On the destination side, tracts with higher numbers of persons in group quarters were associated with increased evacuation inflows ($\beta = 0.023048$), suggesting that institutional or collective living environments may serve as common evacuation destinations. Similarly, higher road density at the destination was positively associated with evacuation flows ($\beta = 0.036927$), reflecting the role of transportation accessibility and network connectivity in facilitating evacuee movement. Additionally, the share of persons age 5 and older who speak English less than well in destination tracts was positively associated with evacuation flows ($\beta = 0.029280$), potentially reflecting the influence of social networks or culturally familiar environments on destination selection.

\section{Discussion}
\label{sec:discussion}

Our study’s use of large-scale mobile device location data offers a unique perspective on the complexity of evacuation behavior. By focusing on the longest stay rather than the first night, we were able to accurately reflect primary evacuation destinations, avoiding the noise typically introduced by transient stops. This method, alongside the integration of social vulnerability, travel impedance, and built-environment characteristics into the modeling framework, allows us to directly evaluate how distance and transportation infrastructure shape evacuation flows and destination choice. Our findings highlight distinct spatial patterns in evacuee movements, revealing the dominance of short-distance relocations under rapid-onset conditions, the constraining role of congestion and network friction at origins, and the facilitating role of accessibility and connectivity at destinations. These results are consistent with prior research emphasizing the importance of incorporating social vulnerability and transportation accessibility into evacuation modeling \citep{Su2020, Sun2024}. To systematically examine these dynamics, we address the following research questions: (1) where evacuees traveled and how far they relocated, (2) how travel impedance, captured through distance and road density, shape evacuation flows, and (3) how social vulnerability influenced evacuation likelihood and destination selection. 

\subsection{Spatial patterns of evacuees during Hurricane Ian}

Spatial patterns in Figures~\ref{fig:OD_flows} and~\ref{fig:spatial} indicate that the majority of evacuation destinations (approximately 92\%) were located within Lee County, suggesting that most individuals relocated only short distances in response to Hurricane Ian. This behavior contrasts with findings from earlier studies, which reported that evacuees often traveled farther than strictly necessary to ensure safety \citep{Dow2002}. The strong concentration of short-distance and intra-county movements also has important implications for model performance and interpretation. Since a large share of OD flows occur over very short distances, the model is heavily influenced by local movements, which tend to be easier to predict due to lower variability and stronger spatial continuity. This may contribute to relatively low error metrics while simultaneously masking greater uncertainty in longer-distance evacuation behavior. Noting this, the model should be interpreted as particularly well-suited for capturing local evacuation dynamics under time-constrained conditions, while predictions for less frequent, long-distance relocations may exhibit higher variability.

Lee County was the primary evacuation destination, with roughly 8\% of evacuees relocated elsewhere, revealing important pat- terns in evacuation behavior. Some traveled to nearby counties such as Collier and Charlotte, which together hosted 2.27\% of evacuees. Others went farther: Miami-Dade and Broward, both over 200 km east, received a combined 1.46\%, while Orange County, approximately 250 km northeast, accounted for another 0.45\%. This indicates that at least 2.27\% of evacuees traveled more than 200 km from home.

These patterns suggest that although proximity dominates overall spatial trends, it is not the sole determinant of evacuation destination choice; evacuees did not always select the geographically closest safe location, indicating that decisions are shaped by a combination of vulnerability, as measured by SVI indicators, and built-environment factors, with several SVI indicators significantly associated with evacuation flows potentially acting as indirect proxies for underlying social networks or support structures.

\subsection{Travel Impendence and the Built Environment}

Our model identifies distance and road density as significant indicators of travel impedance and built-environment conditions that shape evacuation flows. The effects of both variables are strongly conditioned by the rapid-onset nature of Hurricane Ian, which substantially constrained evacuees’ time to prepare, evaluate alternatives, and travel long distances. As a result, evacuation behavior reflects short-distance movements and heightened sensitivity to congestion and accessibility within the transportation network.

\subsubsection{Distance}

Consistent with the spatial patterns found in our data, the estimated coefficient on distance between origin and destination tract centroids was strongly negative ($\beta = -0.241234$), indicating a pronounced distance-decay effect in evacuation flows. This finding aligns with prior evacuation destination choice research, which reports negative distance effects, indicating that the likelihood of selecting a destination declines as travel distance increases, even when safety considerations are salient. For example, \cite{Cheng2011} estimate a statistically significant negative distance coefficient in their evacuation destination choice models, reinforcing distance as a dominant constraint on evacuation destination selection. The much stronger distance effect in our model reflects both the unique context of Hurricane Ian and the fine spatial resolution of our data. Most evacuees relocated only a short distance inland, and this concentration is consistent with cases where warning time is limited and evacuation orders are issued late. Lee County initially is- sued mandatory evacuation orders only around 32 h before the storm’s landfall, and such latency resulted in less preparatory time for taking sufficient protective actions and safe evacuation \citep{Pham2020}. Under such conditions, evacuees prioritized nearby and immediately accessible locations, rather than optimizing for maximum separation from the hazard, underscoring the critical roles of timely emergency alerts and warnings in shaping people’s evacuation decision-making \citep{Kumar2018, Perry2003}.

\subsubsection{Road density}

Road density exhibits asymmetric effects across the origin and destination tracts. On the origin side, higher road density ($\beta = -0.043353$) often serves as a proxy for dense urban cores, where evacuation traffic congestion and operational friction are most pronounced. Prior research shows that these conditions can impede evacuation efficiency and discourage long-distance travel, leading many households to either shelter in place or to undertake shorter, local relocations rather than generating longer origin--destination flows \citep{Dash2007, Staes2021, Rambha2021}. As a result, higher road density at the origin functions as a mobility constraint, amplifying congestion-related barriers under emergency conditions.

In contrast, on the destination side ($\beta = +0.036927$), higher road density more directly reflects accessibility, network connectivity, and infrastructural capacity, including proximity to services, all of which are critical under time-constrained evacuation scenarios. Prior destination choice studies indicate that evacuees tend to favor metropolitan or highly connected areas, especially those with interstate access, because they are easier to reach and offer greater accommodation and service availability \citep{Cheng2008}. When evacuation decisions must be made rapidly, households often have limited opportunity to search for formal accommodations and instead rely on friends, family, and existing social networks to secure temporary shelter, leading evacuation destinations to be spatially constrained by the geographic distribution of social ties and hosting capacity \citep{Fussell2012, Sadri2017}.

In this context, destination tracts with higher road density likely serve as proxies for infrastructure-rich, well-connected areas where evacuees can more readily access housing, services, and social support through established networks. Consistent with this interpretation, \citep{Cheng2008} report positive coefficients for both metropolitan status and interstate proximity in their destination choice models, reinforcing the role of transportation connectivity and access to major corridors in shaping evacuation destination selection when time is limited.

Furthermore, congestion during large-scale evacuations such as Hurricane Irma has been shown to be largely directional, with severe congestion concentrated in evacuating regions rather than in receiving areas \citep{Staes2021, Rambha2021}. Consequently, the congestion-related friction associated with high road density is unlikely to deter evacuees once they reach destination tracts, further reinforcing the observed push–pull asymmetry in road density effects.

\subsection{Social vulnerability and evacuation behavior}

Understanding the factors that attract evacuees during crises is critical for effective emergency management planning. Our findings indicate that, beyond distance and road density, three key social vulnerability factors significantly influenced evacuation flows: vehicle availability, residence in group quarters, and English proficiency.

\subsubsection{Vehicle availability}

Our model produced a negative coefficient ($\beta = -0.029619$) for the percentage of households with no vehicle available per census tract for the origin tract, indicating that fewer households without vehicles, reflecting higher vehicle availability per household, are associated with greater evacuation outflows from the origin tract. This result reflects the well-established role of vehicle availability in shaping household mobility and evacuation readiness \citep{Insani2022, Pham2020}. Access to one or more private vehicles substantially enhances a household's ability to evacuate, influencing not only whether evacuation is feasible, but also when it occurs and how far households can travel, particularly under short evacuation lead times. Greater vehicle availability increases flexibility in departure timing and expands the set of reachable destinations, allowing households to travel longer distances, access a wider range of sheltering options, and select locations offering greater safety or amenities. Because vehicle ownership is strongly correlated with income and economic stability \citep{Klein2017}, zero-vehicle households often face compounded barriers to evacuation, including financial constraints, limited access to temporary lodging, and reliance on constrained public or assisted transportation systems \citep{Elder2007, Willinger2023}. As a result, lower vehicle availability suppresses evacuation demand, leading to fewer observed evacuation trips from origin tracts where private vehicle access is limited.

\subsubsection{Persons living in group quarters}

Our model also identified the count of persons in groups quarters in both the origin tract, ($\beta = +0.026202$) , and in the destination tract ($\beta = +0.023048$), as significant determinants in trip count. The U.S. Census Bureau defines group quarters as organization- managed group living arrangements that provide housing and services, such as medical or custodial care, to restricted populations, including facilities such as college dormitories, nursing homes, group homes, and correctional institutions \citep{CensusBureau2023}.

In the origin tract, this relationship is consistent with prior literature showing that populations residing in group-quarter settings often face distinct evacuation needs and constraints. Institutional facilities such as nursing homes and assisted-living facilities are of- ten subject to mandatory or facility-led evacuations due to medical, mobility, and caregiving requirements, which can generate elevated outbound evacuation flows from origin tracts \citep{Dostal2015, Smith2009}. Also, public shelters and institutional facilities may lack accessible infrastructure, adequate privacy, child-friendly accommodations, or specialized medical and dietary re- sources required by residents with physical, cognitive, or health-related vulnerabilities. As \citep{Phraknoi2023} note, emergency re- lief systems frequently rely on standardized supplies and collective living arrangements that are not tailored to the needs of older adults or individuals requiring specialized care. These structural limitations can make sheltering in place or remaining within institutional facilities less viable during hazard events, increasing the likelihood of relocation \citep{Smith2009}. In such cases, evacuees often rely on informal social networks, including family members, friends, or caregivers, as safer and more flexible alternatives to institutional sheltering, despite distance considerations \citep{Pham2020}.

On the destination side, higher concentrations of group-quarter facilities may signal the presence of supportive infrastructure and social capacity that are especially critical during evacuations. Populations served by these facilities, particularly older adults and individuals with disabilities, experience heightened physical, mobility, and caregiving challenges, making sheltering in place riskier and increasing the need for early or assisted relocation during hazard events \citep{Dostal2015, Bayraktar2018}. Destination tracts with more group-quarter facilities are therefore more likely to offer accessible transportation, medical services, trained staff, and specialized sheltering capacity. Prior studies show that evacuees frequently relocate to destinations characterized by institutional housing, shelters, hotels, or the homes of family and friends, which tend to cluster spatially in areas with existing group-quarter infra- structure \citep{Hasan2011, Sadri2014, Thakur2022}. Taken together, the positive destination-side coefficient suggests that tracts with group-quarter facilities function as focal points, where service capacity, accessibility, and institutional readiness are spatially concentrated, enhancing their ability to receive evacuees with greater support needs \citep{Flanagan2011, Lindell2011}.

\subsubsection{English proficiency}

Finally, our model also identified English proficiency as a significant variable in the destination tract. The coefficient ($\beta = +0.029280$) implies that tracts with a higher percentage of persons aged 5 and older who report speaking English less than well attract more evacuees and experience higher trip volumes. This relationship likely reflects the role of social networks and co-linguistic community structures in shaping evacuation destination choices. Prior research shows that households with limited English proficiency often rely more heavily on informal social networks, such as family, friends, and co-ethnic communities, during disaster response due to barriers in accessing official information, warnings, and services \citep{Fussell2010, Reininger2013}.As a result, evacuees may preferentially relocate to destinations where linguistic similarity and community familiarity reduce communication barriers and uncertainty.

Empirical studies of post-disaster mobility indicate that evacuation destinations are often shaped by existing social and ethnic net- works, rather than by proximity alone. When relocation occurs through family or community ties instead of formal sheltering systems, evacuees tend to concentrate in areas where immigrant or co-ethnic communities are already established \citep{Fussell2012, Vu2009, Pais2008}. In this light, higher destination-side concentrations of limited English proficiency likely capture the presence of linguistically accessible environments—places where evacuees can draw on social support, informal caregiving, translation assistance, and culturally familiar resources. More broadly, research shows that linguistic isolation influences both evacuation decision-making and post-disaster mobility, shaping not only whether households evacuate, but also the destinations they select \citep{Elder2007, Uekusa2019}.

\subsection{Policy implications and limitations}

Our findings have important policy implications for improving hurricane evacuation strategies and strengthening community resilience. By integrating the Social Vulnerability Index (SVI) into the evacuation model, this study underscores the need to prioritize support for socially vulnerable populations, including individuals living in group quarters, households without access to private vehicles, and communities with limited English proficiency. Emergency decision-makers should allocate resources equitably to ensure that evacuation options are both safe and accessible for these populations. Key policy actions include expanding transportation ser- vices for evacuees without vehicles, enhancing multilingual outreach and communication, and equipping shelters with appropriate resources for individuals with disabilities or specialized care needs. Strengthening these support systems within socially vulnerable communities can improve evacuation effectiveness and help reduce inequities in disaster response.

However, this study has limitations. While mobile device location data provides valuable insights, it may not capture all evacuees, particularly those without mobile access or who disable location services. Also, GPS records do not include individual sociodemographic information, making it difficult to determine which populations are represented in the dataset and to assess potential biases. As a result, the data may not be fully representative of the broader population and may systematically underrepresent certain groups, such as older adults, lower-income households, minorities, or individuals with limited access to smartphones \citep{PewResearch2025, Wang2018, Jiang2013, Li2024}. In addition, spatial and temporal coverage may vary due to differences in device usage, signal availability, and data collection processes, which can further contribute to uneven representation across regions \citep{Wang2018, Jiang2013}. Although SVI data provide a useful proxy for vulnerability, they are aggregated at the census tract level and may contain sampling errors or overlook localized variations, potentially obscuring within-tract differences.

Furthermore, the term “group quarters” cannot be further defined than just its broad definition by the U.S. Census data. Because the dataset does not provide group-quarter subtype information at the tract level, we are unable to distinguish which specific types of group-quarter facilities are driving the observed positive coefficients in the origin and destination tracts. As a result, our interpretation relies on inference informed by prior evacuation and disaster-response literature rather than direct observation. Because we can- not directly identify whether evacuees are relocating to emergency shelters, medical facilities, or other institutional settings, the positive destination-side coefficient should be interpreted as reflecting general institutional capacity and collective housing availability, rather than any single destination type. Similarly, the positive origin-side coefficient may reflect a combination of proactive institutional evacuations, limited shelter-in-place feasibility, and heightened vulnerability among populations residing in group quarters, rather than uniform behavior across all facility types.

Additionally, our study does not explicitly capture psychological or perceptual factors that influence evacuation behavior. Prior research shows that decisions such as route choice or departure timing often result from a series of internal cognitive steps shaped by risk perception, memory of past events, familiarity with road networks, subjective evaluations of congestion or danger, and personal comfort with uncertainty \citep{Huang2015}. These perceptual processes can lead evacuees to select routes or destinations that feel safer or more efficient even when they differ from objectively optimal choices. While our direct demand model indirectly reflects some of these influences through variables such as distance and vehicle availability, it does not explicitly model the psychological mechanisms that guide how evacuees interpret information and navigate during emergencies. Future work incorporating these behavioral dimensions would provide a more holistic understanding of evacuation dynamics.

Finally, findings from Lee County may not be generalizable to regions with different demographics or geographies, highlighting the need for future research in diverse locations. Evacuation behavior is highly context-dependent, and factors such as regional housing characteristics, transportation infrastructure, social networks, and cultural norms can meaningfully alter both destination preferences and mobility patterns. For example, rural areas may require longer-distance evacuations due to limited nearby options, while dense urban regions may experience more constrained movement due to congestion. These differences reflect the spatial nature of evacuation flows, in which nearby locations and similar regions often exhibit correlated behavior. This also highlights a methodological limitation of using 10-fold cross-validation. Because this approach relies on random partitioning of OD pairs, it does not explicitly account for spatial dependence. Consequently, geographically similar observations may appear in both training and testing folds, leading to slightly optimistic performance estimates that reflect generalization within similar spatial contexts rather than across dis- tinct regions. Future work could address these limitations by implementing spatially structured validation strategies, such as geographically grouped folds, and by applying this framework across varied geographic settings to better assess robustness and improve its applicability to broader disaster scenarios.

\section{Conclusion}
\label{sec:conclusion}

Overall, this study has produced vital insights that can inform more effective evacuation strategies, better resource allocation, and improve safety and resilience for populations during natural disasters. Implementing a multiplicative model combined with large-scale mobile device location data, we were able to predict evacuation patterns while emphasizing the role of social vulnerability. Our findings indicate that evacuation outcomes are strongly influenced by transportation access, housing and institutional characteristics, and destination-side infrastructure. Expanding shelter capacities, multilingual evacuation notices and efforts, and improving transportation connectivity (roads and public transportation) can help ensure that evacuees relocate to safer and more resilient locations.

Future research can build on this work in several ways. First, integrating survey data, interviews, and collaborations with local emergency management agencies would help validate model outputs and provide deeper insight into the motivations underlying observed evacuation patterns. Second, incorporating additional factors, such as psychological responses, cultural influences, health conditions, and caregiving responsibilities, could further refine the model. Finally, applying this framework across different geographic regions, hazard types, and disaster contexts would enhance its generalizability and support the development of more inclusive and equitable evacuation strategies.

\section*{CRediT Authorship Contribution Statement}
\textbf{Alessandra Recalde:} Writing -- review \& editing, Writing -- original draft, Visualization, Validation, Methodology, Investigation, Formal analysis.
\textbf{Luyu Liu:} Writing -- review \& editing, Supervision, Methodology, Data curation, Conceptualization.
\textbf{Xiaojian Zhang:} Writing -- review \& editing, Writing -- original draft, Supervision, Methodology, Data curation.
\textbf{Sangung Park:} Writing -- review \& editing, Validation, Supervision.
\textbf{Shangkun Jiang:} Writing -- review \& editing, Data curation.
\textbf{Xilei Zhao:} Writing -- review \& editing, Validation, Supervision, Project administration, Methodology, Funding acquisition, Data curation, Conceptualization.

\section*{Declaration of Competing Interest}
The authors declare that they have no known competing financial interests or personal relationships that could have appeared to influence the work reported in this paper.

\section*{Acknowledgment}
This research was supported by the U.S. Department of Transportation (USDOT) through the Center for Transient-oriented Communities (CETOC) (Grant No.\ 69A3552348337). This material is also based upon work supported by the National Science Foundation (NSF) under Grant Nos.\ 2338959 and 2303578, and by an Early-Career Research Fellowship from the Gulf Research Program of the National Academies of Sciences, Engineering, and Medicine. Any opinions, findings, and conclusions or recommendations expressed in this material are those of the authors and do not necessarily reflect the views of USDOT, NSF, or the Gulf Research Program of the National Academies of Sciences, Engineering, and Medicine. The authors acknowledge using ChatGPT to check grammar errors and improve language.

\bibliographystyle{apalike}
\bibliography{references}

\end{document}